\documentclass[twocolumn,showpacs,preprintnumbers,amsmath,amssymb]{revtex4}
\usepackage{graphicx}
\usepackage{dcolumn}
\usepackage{bm}
\newcommand{\gsim}{\lower.7ex\hbox{$\;\stackrel{\textstyle>}{\sim}\;$}}
\newcommand{\lsim}{\lower.7ex\hbox{$\;\stackrel{\textstyle<}{\sim}\;$}}

\newcommand{\cm}{~\mathrm{cm}}

\newcommand{\mX}{m_{\scriptscriptstyle \chi}}

\newcommand{\fout}{f_{\rm out}}

\newcommand{\TeV}{\,\mathrm{TeV}}
\newcommand{\GeV}{\,\mathrm{GeV}}

\newcommand{\be}{\begin{equation}}
\newcommand{\ee}{\end{equation}}
\newcommand{\bea}{\begin{eqnarray}}
\newcommand{\eea}{\end{eqnarray}}

\newcommand{\bef}{\begin{figure}[htbp]\begin{center}}
\newcommand{\eef}{\end{center}\end{figure}}

\begin{document}
\preprint{
SU-ITP-09/44}
\title{
High Energy Electron Signals from Dark Matter Annihilation in the Sun}

\author{Philip Schuster$^1$, Natalia Toro$^2$, Neal Weiner$^3$, Itay
  Yavin$^3$} 
\affiliation{
$^1$ Theory Group, SLAC National Accelerator Laboratory,
 Menlo Park, CA 94025, USA\\ 
$^2$ Stanford Institute for Theoretical Physics, Stanford University, Stanford, CA 94305, USA \\
$^3$ Center for Cosmology and Particle Physics, Department of Physics, \\NewYork University, NewYork, NY10003, USA
}%

\date{\today}

\begin{abstract}
In this paper we discuss two mechanisms by which high energy electrons resulting from dark matter annihilations in or near the Sun can arrive at the Earth. Specifically, electrons can escape the sun if DM annihilates into long-lived states, or if dark matter scatters inelastically, which would leave a halo of dark matter outside of the sun. Such a localized source of electrons may affect the spectra observed by experiments with narrower fields of view oriented towards the sun, such as ATIC, differently from those with larger fields of view such as Fermi. We suggest a simple test of these possibilities with existing Fermi data that is more sensitive than limits from final state radiation. If observed, such a signal will constitute an unequivocal signature of dark matter.
\end{abstract}
\pacs{12.60.Jv, 12.60.Cn, 12.60.Fr}
\maketitle

\section{Dark Matter in the Sun}

High-energy particles from dark matter (DM) capture and annihilation
in the Sun offer a striking signature of dark matter
\cite{Press:1985ug,Silk:1985ax}.  The study of energetic neutrinos from the Sun~\cite{Krauss:1985aaa,Gaisser:1986ha,Griest:1986yu}
has received great attention in this context, as it is assumed that
charged products would not escape the Sun's interior. Recent data and
theoretical developments call this assumption into question. In
particular, the solar signatures of dark matter annihilation in the
Sun can be greatly altered for dark matter that annihilates into a new
force carrier
~\cite{Finkbeiner:2007kk,Pospelov:2007mp,ArkaniHamed:2008qn}, or for
inelastically interacting dark matter
(iDM)~\cite{TuckerSmith:2001hy}. In this paper, we discuss how either
scenario allows charged particles from DM annihilations in the Sun to
reach the Earth, and the observational signatures of this effect.

\begin{figure}
\includegraphics[width=0.4\textwidth]{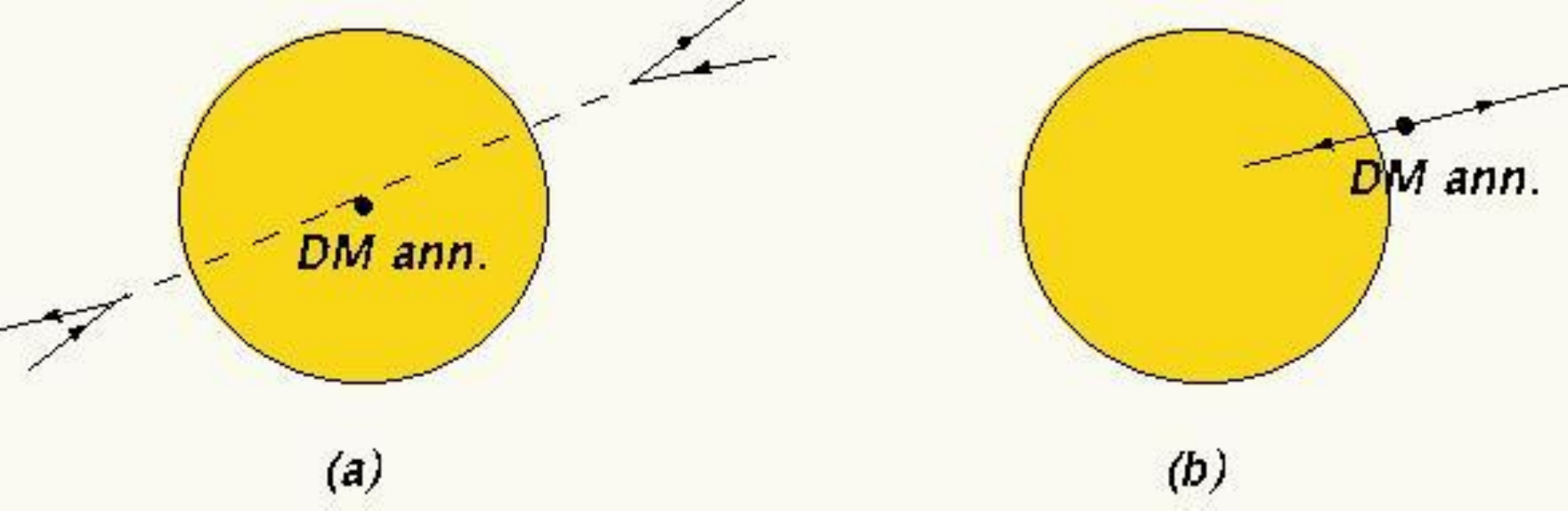}
\caption{Two possible escape mechanisms for high energy charged particles from DM annihilations in the Sun. (a) DM may annihilate into long-lived states which first escape the Sun and only later decay. (b) DM may annihilate outside the Sun. \label{fig:escape}}
\end{figure}

In the first case, illustrated in Fig.~\ref{fig:escape}(a), DM
annihilates into long-lived particles, such as scalars associated
with a new gauge sector. These long-lived particles can easily escape
the Sun, and their subsequent decay in the solar system into
electrons, muons, or charged pions can be detected.  In the second
case, DM captured through inelastic scattering may lack the minimum
kinetic energy required to scatter again.  If the elastic scattering
cross-section is small, DM forms a loosely bound halo around the Sun
and can annihilate outside the Sun as shown in
Fig.~\ref{fig:escape}(b).

In either scenario, satellite observatories such as Fermi~\cite{Abdo:2009zk} can detect
the electronic annihilation products as a cosmic ray electron excess
strongly correlated with the Sun's direction. If observed, such an
effect is an unequivocal signature of DM since no known astrophysical
phenomena can generate such a high-energy electron flux from the
Sun. This type of signature may offer a unique probe of inelastically
interacting dark matter, for which direct detection constraints are
quite weak.

Our estimates will show that a solar flux $F\sim 10^{-4}~{\rm m}^{-2}{\rm s}^{-1}$ of particles above several hundred GeV should be detectable by experiments such as Fermi.  Thus, only a small fraction of DM captured in the sun must annihilate through these channels to observe an effect.  Indeed, if for a given DM mass we take the largest cross-section allowed by direct detection limits on spin-independent elastic scattering  ($\sigma_{\rm SI}\approx 0.5~ (3) \times 10^{-43} \cm^2 $ for $\mX\approx 0.1~ (1) \TeV$)~\cite{Angle:2007uj, Ahmed:2008eu}, then DM is captured at a rate~\footnote{This formula gives an excellent approximation to the capture obtained from the more precise computation found in~\cite{Gould:1987ir, Gondolo:2004sc}.},

\begin{equation}
C_{\odot} \approx 1.4 \times10^{21} \mbox{ s}^{-1}\left(\frac{\TeV}{\mX} \right)^{2/3}.
\end{equation}

IDM models allow much larger
cross-sections $\sigma_n \gtrsim 10^{-40} \cm^2$ and hence considerably higher capture rates~\cite{Nussinov:2009ft,Menon:2009qj}. For cross-sections in this range, the DM density accumulated over the age of the Sun is high enough that DM capture and annihilation rate ($\Gamma_A$) reach equilibrium so that $\Gamma_A = \frac{1}{2} C_\odot$. Assuming one observable product per annihilation actually leaves the Sun, the flux at the Earth is
\bea
F\sim && 5\times 10^{-3}~{\rm m}^{-2}{\rm s}^{-1} \mbox{(elastic)} \label{eqn:flux1} \\
F\sim && 50~ {\rm m}^{-2}{\rm s}^{-1} \mbox{(inelastic).} \label{eqn:flux2}
\eea
Both estimates are significantly larger than the sensitivity limit, so
that even very sub-dominant reactions of the form discussed here can
be probed by careful analysis of electronic cosmic ray
data. Interestingly, a signal as large as \eqref{eqn:flux1} would
constitute a significant fraction $\sim 10\%$ of the electronic cosmic
rays above $300$ GeV observed by PPB-BETS~\cite{Torii:2008xu}, ATIC~\cite{2008zzr}, and Fermi, and could
lead to an observable feature in the overall signal rate. This is
particularly important when considering balloon-based experiments
situated at the south pole, such as ATIC and PPB-BETS, because
spectral sculpting can be large for signals with a sharp
directionality.

\section{Escape Mechanisms for Charged Particles from Solar DM Annihilation}\label{sec:EscapeMechanisms}

\subsection{Long-Lived States}
We first consider the case as initially proposed in \cite{Finkbeiner:2007kk} where DM annihilates into the SM via some intermediate states that then decay to Standard Model particles through a small mixing. This is the case, for example, if DM interacts through some new force, as has been employed to explain astrophysical anomalies \cite{Finkbeiner:2007kk,Pospelov:2007mp,Cholis:2008vb,ArkaniHamed:2008qn,Pospelov:2008jd,Cholis:2008qq}. These new bosons can decay back into the SM or cascade further in the dark sector. Some of the intermediate states may be sufficiently long-lived that they escape the Sun before decaying into SM states~\cite{Baumgart:2009tn,Batell:2009yf,Essig:2009nc} (see \cite{Schuster:2009au,Bjorken:2009mm} for existing constraints on such long-lived states). As suggested in \cite{Jedamzik:1999di,Jedamzik:2007cp}, Lithium abundance discrepancies \cite{VangioniFlam:1998gq, Asplund:2005yt, Cyburt:2008kw} may be resolved by late-decaying ($\tau \sim 10^3$s) particles of weak-scale mass. This offers another motivation, as DM annihilations into such an intermediate state would allow the eventual decay products to escape the Sun. 

\begin{figure}
\includegraphics[width=0.4\textwidth]{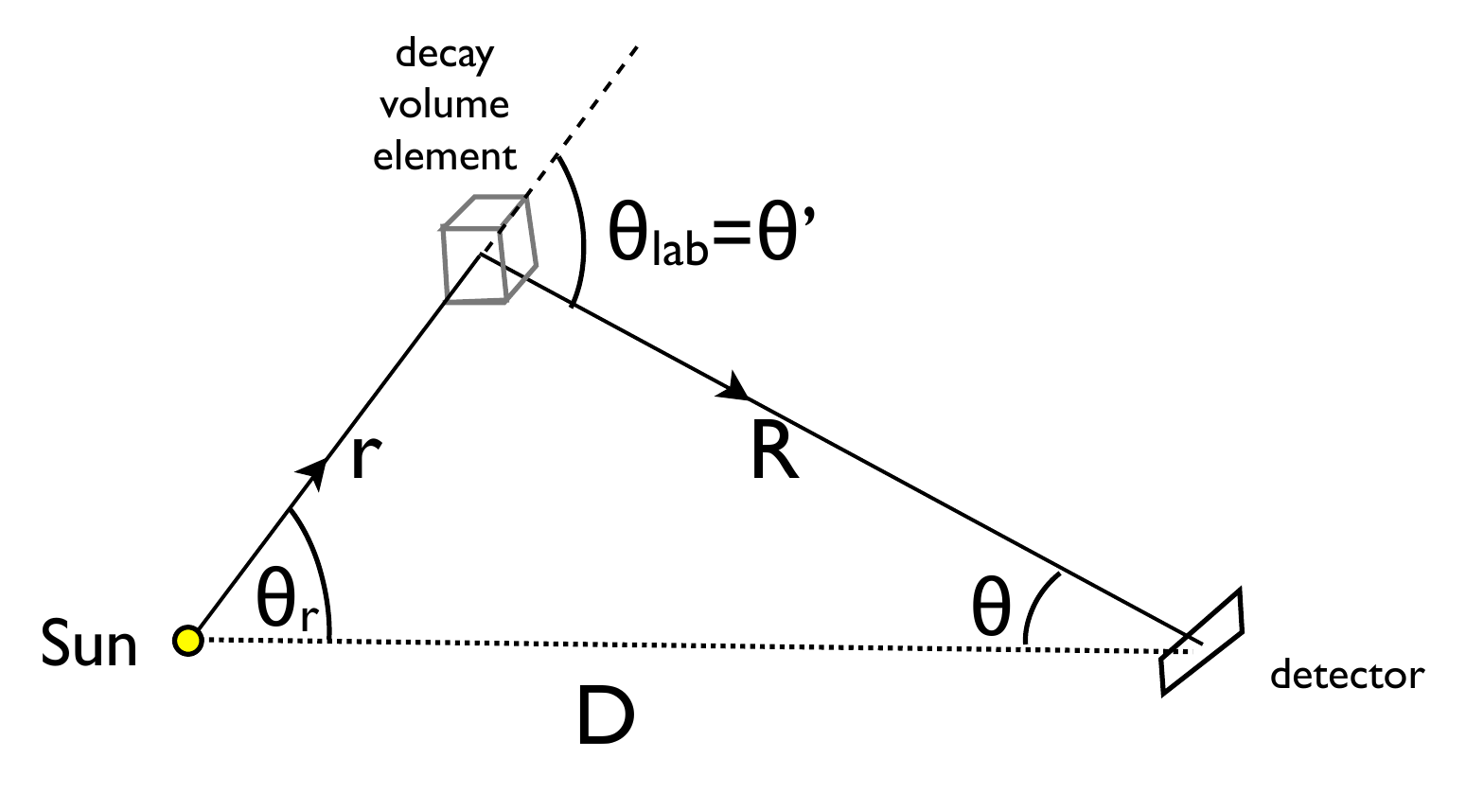}
\caption{Geometry of DM annihilations into long-lived states, followed by 2-body decays into charged particles. $\theta_{lab}$ is the laboratory angle of the forward going decay product.}
\label{fig:Geometry}
\end{figure}

Fig.~\ref{fig:Geometry} illustrates the geometry of our setup and defines notation. 
To keep the discussion somewhat general, we calculate the profile of decay products assuming that DM annihilates in the center of the Sun into a 
particle with proper lifetime $\tau$, mass $m_I$, and velocity $\beta$. For simplicity, we assume that the intermediate particle decays via a 2-body reaction into an electron-positron pair.  From Fig.~\ref{fig:Geometry}, we see that the rate per unit area, per unit solid angle, per unit energy of electrons observed by a detector is,
\bea
&& \frac{d\dot{N}_{det}}{d\cos{\theta} dAdE} = \int dR R^2  ( \frac{C_\odot e^{-r/\gamma c\beta\tau}}{4\pi r^2\left( \gamma c\beta\tau\right) } )\nonumber \\
&& \times \frac{d\Gamma}{d\cos{\theta_{cm}}} \frac{d\cos{\theta_{cm}}}{d\cos{\theta'}} \frac{\delta(E-E(\theta,D,R))}{R^2 },
\eea
where $\theta_{cm}$ and $\theta'$ are related via boosting from the CM frame to the solar frame, $\frac{d\Gamma}{d\cos{\theta_{cm}}}$ is the 2-body decay distribution of the intermediate state given in its rest frame, and all quantities are evaluated in the geometry of Fig.~\ref{fig:Geometry}. 
The energy delta function is simply enforcing that the solar frame energy is the energy of the boosted decay products. In particular,
\be
E(\theta,D,R) = \gamma m_I \frac{1}{2} (1 + \beta \cos{\theta_{cm}}(\theta,R,D) ).
\ee

Integrating over $R$ and $\phi$ can be done using the delta function and assuming azimuthal symmetry. We treat $\gamma$, $\beta$ and $D$ as fixed. The delta function localizes $R$ as a function of $E$ (where $E$ is the energy of the observed lepton), and $\theta$. Also note that on the support of the delta function, $\theta'$ is a function of $E$ alone, and $r$ is a function of $E$ and $\theta$.
Introducing the dimensionless energy variable,
\be
x=\frac{2E}{(1+\beta)\gamma m_I}, 
\ee
we have that the minimum and maximum value for $x$ is,
\be
x_{min}=\frac{1-\beta}{1+\beta} \leq x \leq 1.
\ee
On the support of the delta function we have, 
\bea
\cos{\theta'} &&= \frac{1}{\beta}(1-\frac{1-\beta}{x}) \\
\sin{\theta'} && = \frac{1}{\beta\gamma x}\sqrt{(1-x)(x-x_{min})} \\
r &&= \frac{D\gamma \beta x \sin{\theta}}{\sqrt{(1-x)(x-x_{min})}},
\eea 
 where the last equality follows simply from the law of sines: $r=\frac{D\sin{\theta}}{\sin{\theta'}}$.

Putting all the pieces together, and using $\frac{d\Gamma}{d\cos{\theta_{cm}}}=1/2$ for isotropic 2-body decay, we have,
\bea
&& \frac{d\dot{N}_{det}}{d\cos{\theta}dAdx} = \frac{C_\odot}{4\pi L D} \frac{1+\beta}{2\beta \sin{\theta}} e^{-r/L} \nonumber \\
&&  \times\frac{\gamma \beta x}{\sqrt{(1-x)(x-x_{min})}} \Theta(x_{max}-x),
\eea
where $x_{max}=\frac{1-\beta}{1-\beta\cos{\theta}}$, $x_{min} = (1-\beta)/(1+\beta) $, and $L\equiv\gamma c\beta \tau$.

\begin{figure}
\includegraphics[width=0.4\textwidth]{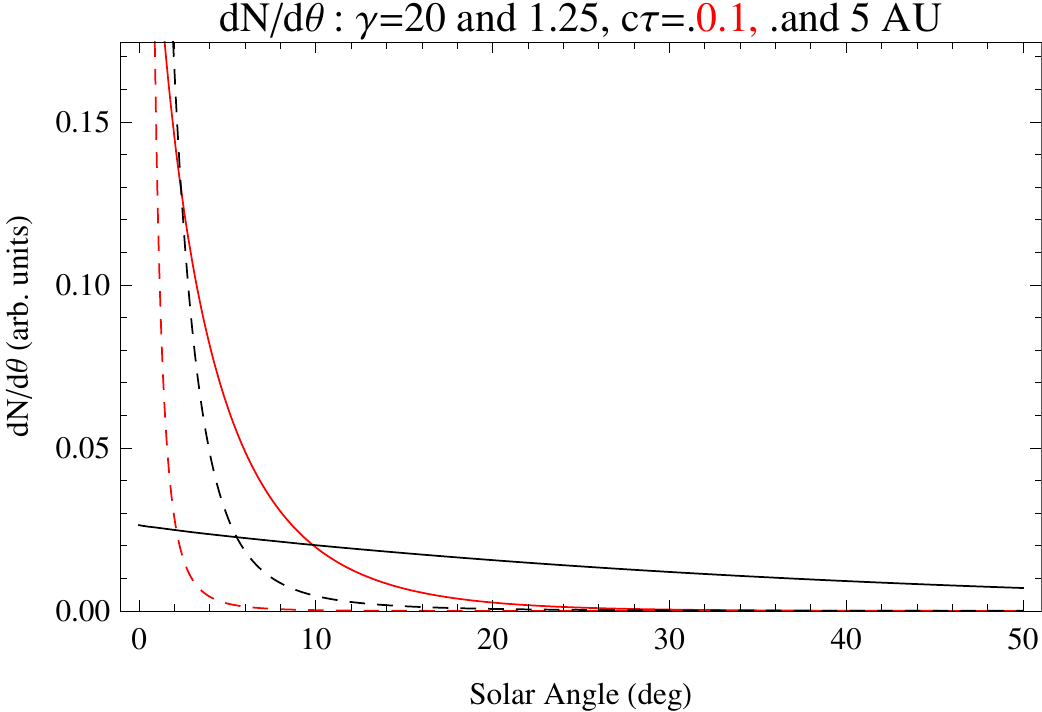}
\caption{Angular distribution from a single 2-body cascade, in degrees, for various decay lengths and long-lived particle velocities.  We consider $\gamma=1.25$  (red) and 20 (blue), and $\gamma c\tau=0.1$ (dashed) and $5$ (solid) AU.  If dark matter annihilates through a new force at the GeV-scale, even larger $\gamma$ are expected.  Both decay length $\gtrsim 1$ AU and moderate $\gamma$ are required to broaden the electron and FSR photon fluxes significantly.}
\label{fig:profile}
\end{figure}

For $\beta\approx1$, and $\gamma c \tau \lsim \mbox{AU}$, the decay products travel in approximately straight lines from the Sun, and so the profile is obviously peaked strongly at the Solar center relative to the detector. For $\beta\lsim 1$ and $\gamma c \beta\tau \gsim \mbox{AU}$, the directionality relative to the Sun is significantly broadened --- see Fig.~\ref{fig:profile}.

\subsection{DM Annihilations Outside the Sun}
Another mechanism allowing electronic annihilation products to reach
the Earth is for DM to annihilate near the surface of the Sun.  
In the standard scenario of elastic scattering, DM undergoes many further collisions with matter in the Sun after being captured. If the spin-independent elastic scattering cross-section is larger than about $10^{-47}~{\rm cm^2}$ the collision rate is sufficiently high that the captured DM will quickly thermalize with the rest of the matter in the Sun and concentrate in the inner core~\cite{Griest:1986yu} (this estimate is based on the iron density and abundance in the sun). We therefore typically do not expect appreciable annihilation rate outside the Sun, as was clearly demonstrated in Ref.~\cite{Sivertsson:2009nx}.

The situation is quite different in the iDM scenario. In particular, DM particles scatter only a few times before they lose enough energy to render further scattering kinematically forbidden. If the elastic scattering component is much smaller than $10^{-47}{\rm cm^2}$, the DM will never reach thermal equilibrium with matter in the Sun resulting in only mild core concentration \footnote{Such small elastic cross-sections are expected in cases where the inelasticity is achieved by lifting the degeneracy between the two CP eigenstates of a scalar or the two Majorana components of a Dirac fermion.}. 
A non-negligible fraction of captured DM is then bound in elliptical orbits of order the size of the Sun and can then annihilate outside of the Sun. This is different from the usual case of elastic scattering, which was carefully investigated in Ref.~\cite{Sivertsson:2009nx}  in that the WIMPs can spend a very long (of the order of the lifetime of the Sun) time outside the Sun since they \textsl{never} lose enough energy to fall into the center and get trapped.  

The iDM capture rate in the Sun was calculated in \cite{Nussinov:2009ft,Menon:2009qj}. While the DM no longer thermalizes with matter in the Sun, the annihilation rate does track the capture rate since the WIMPs will continue to accumulate until equilibrium is reached. Using $\sigma_n = 10^{-40}\cm^2$, the annihilation rate outside the Sun is then related to the capture rate as,
\begin{eqnarray}
\label{eqn:GammaAout}
\Gamma_A^{\rm out} &=& \frac{1}{2} C \fout  \\ \nonumber &=& 1.5 \times 10^{23} ~\fout\left(\frac{\TeV}{\mX}\right)^2 \left(\frac{\sigma_n}{10^{-40}{\rm cm^2} } \right)~{\rm s^{-1}} 
\end{eqnarray}
where a conservative estimate\footnote{The quantity $\fout$ was calculated based on the calculations presented in Ref.~\cite{Nussinov:2009ft}. An analytical solution for the trajectories inside and outside the Sun was used in a Monte-Carlo simulation of the capture and entrapment process. The simulation results, which include the final trajectory of a large number of particles, were used to calculate the asymptotic density for a give WIMP mass, $\mX$ and inelasticity $\delta$.} of $\fout$ is,
\begin{equation}
\fout =\frac{\int_{r_{Sun}}^\infty n_{DM}(r)^2 d^3r}{ \int_0^\infty n_{DM}(r)^2 d^3r}.
\end{equation}
The resulting flux of charged particles at the Earth is,
\be
F_{earth}\approx  \fout\left(\frac{\TeV}{\mX}\right)^2 \left(\frac{\sigma_n}{10^{-40}{\rm cm^2} } \right)~{\rm m}^{-2}{\rm s}^{-1}. \label{eq:surface}
\ee

\begin{figure}
\includegraphics[width=0.4\textwidth]{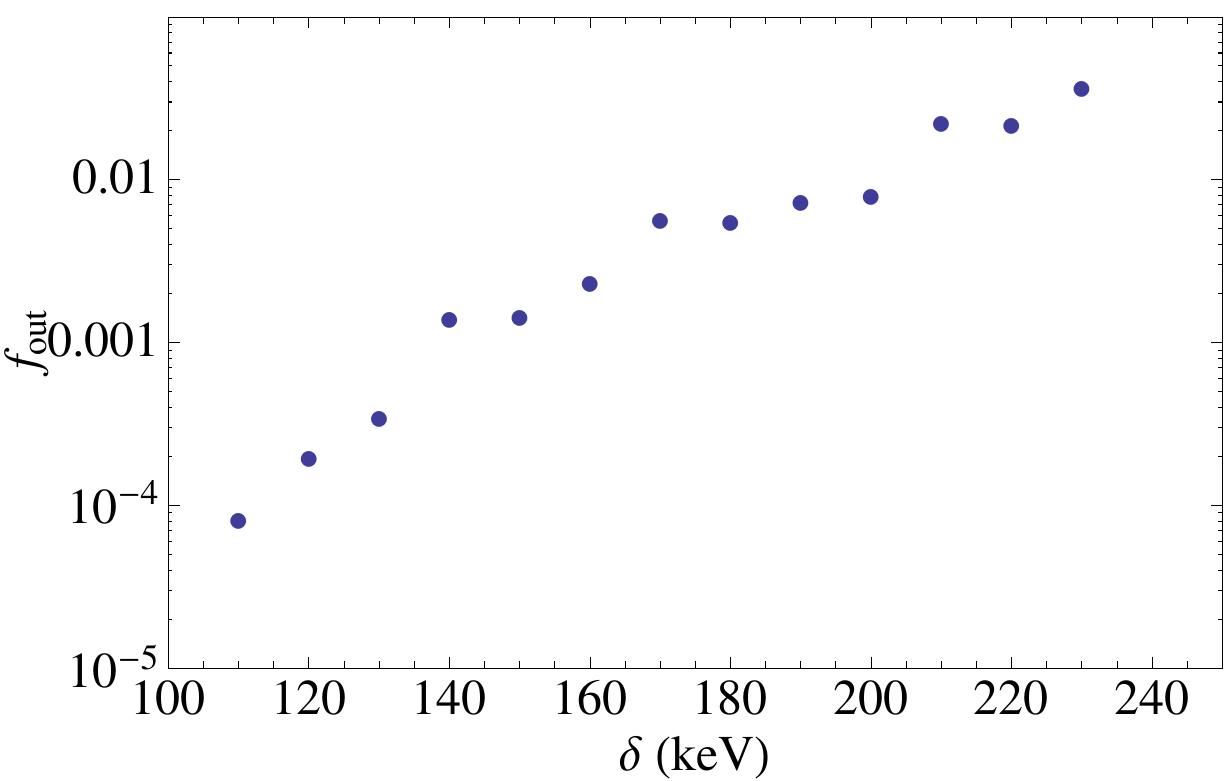}
\caption{A plot of $\fout$ against the inelasticity parameter $\delta$ for $\mX=\TeV$. The dependence on $\mX$ is weak since the dynamics is determined only by the reduced mass of the DM particle-nucleus system. For much higher values of the inelasticity, the capture rate rapidly drops as it becomes impossible for the DM particles to scatter anywhere in the Sun.}
\label{fig:fout}
\end{figure}

To compute $\fout$ (shown in Fig.~\ref{fig:fout}), we simulated the accumulation of DM in the Sun assuming only inelastic collisions as described in Ref.~\cite{Nussinov:2009ft}. 
Notice that even at low inelastic thresholds with $\fout\sim 10^{-4}$, the annihilation rate outside the Sun (eq. \ref{eqn:GammaAout}) and resulting charged particle flux (eq. \ref{eq:surface}) can be quite large --- much larger than our estimated Fermi's sensitivity. 

\section{Existing Constraints and Signatures}\label{sec:Constraints}

\subsection{FSR Constraints}
Any electronic production near the Sun or along our line of sight also
yields $\gamma$-rays from final state radiation (FSR).  Therefore,
these processes are bounded by Fermi's measurement of the $\gamma$-ray
spectrum from the Sun up to $\sim 10\GeV$~\cite{Giglietto}.  
For illustration we consider DM annihilation $\chi\chi\rightarrow \phi_2\phi_2$ with two possible decays for $\phi_2$: one-step decay $\phi_2 \rightarrow e^+e^-$; and two-step decay $\phi_2\rightarrow \phi_1\phi_1$ followed by $\phi_1\rightarrow e^+e^-$. The
limits have been obtained in \cite{Bjorken:2009mm, Batell:2009zp, Schuster:2009au} using the
simplified expressions for differential photon flux from
\cite{Mardon:2009rc}.  Comparable limits can be obtained from the
Milagro search for high-energy gamma rays, which is sensitive to very
high-energy ($>100 \GeV$)~\cite{Atkins:2004qr}. These constraints consequently severely restrict high-energy electronic emissions from the Sun. For example, an injection spectrum normalized to
the positron flux observed by PAMELA would produce a $\gamma$-ray flux
at 5--10 GeV two orders of magnitude larger than observed. It is
therefore difficult to explain the low energy positronic excess as
coming from annihilations of DM in the Sun without running into a
serious conflict with the Fermi $\gamma$-ray observations.  Likewise,
attempting to explain the discrepancy between ATIC and Fermi data by
normalizing the electron flux to the difference in flux at higher
energies $E\sim500\GeV$, the resulting $\gamma$-ray flux is still
about an order of magnitude too large. FSR constraints generically restrict the Solar electronic flux to be at most $\mathcal{O}(10\%) $ of the total flux observed by Fermi.  Such a rate is still significantly above the expected sensitivity of a dedicated search for high-energy electrons from the Sun.  Moreover, much larger electronic fluxes are possible if $m_{\phi_1} \approx 2m_e$, in which case FSR is phase-space suppressed, or if the injection profile is spatially broad, as is the case for $\beta\lsim 1$, $\gamma c \tau \gsim \mbox{AU}$ (see Fig.~\ref{fig:profile}).

\subsection{Electronic Signatures}
Testing the scenarios described above is possible with existing Fermi electron data \cite{Abdo:2009zk}. One possibility is to form an asymmetry variable based on the differential flux coming from outside of the sphere defined by the Sun (night) or from the inside (day). One can then eliminate much of the systematics and be sensitive to any residual flux from the general direction of the Sun. 
\begin{equation}
A_{\rm dn} = \left. d\Phi^e/dE\right|_{\rm day} -\left. d\Phi^e/dE\right|_{\rm night}
\end{equation}
Before discussing some more precise signatures, we briefly consider the trajectory of an electron emitted from the sun. The strong magnetic fields in the Sun's vicinity~\cite{Babcock1955ApJ} will affect any electronic flux emanating from the Sun. Since the gyromagnetic radius is smaller than the curvature in the fields the electrons will follow the field lines, causing them to diffuse and deflect. Very close to the Sun's surface, an approximate dipole field of $1 \ {\rm gauss}$ is present which will result in a diffusion of the isotropic flux. Away from the Sun's surface, the well known Parker spirals~\cite{Parker1958ApJ} will also alter the electrons' trajectories causing them to arrive at the Earth from an apparent source shifted by up to $30^o$ relative to the Sun's position. These two effects mean that we must consider a few possibilities when searching for electrons from the sun. In the presence of a positive signal, these effects will clearly need to be better understood, as well as the effects of the Earth's magnetic field on the acceptance of experiments situated at the poles, such as ATIC and PPB-BETS, in order to interpret any result.

\begin{figure}
\includegraphics[width=0.4\textwidth]{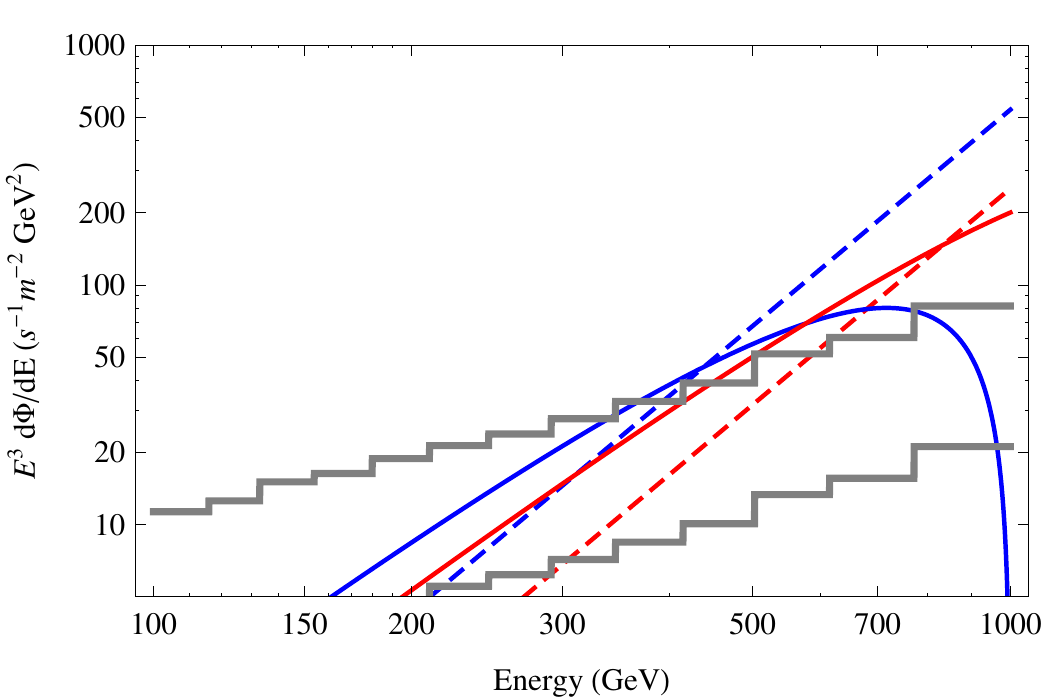}
\caption{Electronic energy spectrum for a one-step (dashed) and two-step (solid) decays for a DM mass of 1 TeV (red) and 2 TeV (blue). These correspond to the process $\chi\chi\rightarrow \phi_2\phi_2$ followed by $\phi_2 \rightarrow e^+e^-$ (one-step) or $\phi_2\rightarrow \phi_1\phi_1$ and $\phi_1\rightarrow e^+e^-$ (two-step).  The curves have been normalized to avoid FSR limits as discussed in the text. Also plotted are estimates of Fermi's statistical sensitivity for half sky (upper) and $30^{o}$ searches, given the statistics of~\cite{Abdo:2009zk}}
\label{fig:FERMIsensitivity}
\end{figure}

However, there are simple and straightforward approaches that can increase sensitivity without a detailed modeling of the signal. In particular, we can consider the more precise signature of the differential flux, $d\Phi_{\rm Sun}/dE(\Delta\theta)$ in a circle of size $\Delta\theta$ centered on the Sun. By comparing to a region of the sky of the same size but an angular distance $\bar{\theta}$ away (a ``fake Sun'') , an estimate of the expected background flux can be obtained. (Optimally the fake sun should be in the hemisphere opposite that of the real sun.)
If $\Phi_{\rm fake}$ is the flux through the ``fake Sun'', then a useful search variable is the asymmetry,
\begin{equation}
A_{\theta} = d\Phi_{\rm Sun}(\Delta\theta)/dE - d\Phi_{\rm fake}(\Delta\theta)/dE
\end{equation}
studied as a function of energy. We expect that the background rates
will be well approximated by the ``fake Sun'' for regions of size
$\Delta\theta\gsim 10$ degrees at energies above $\sim 100 \GeV$. For
smaller circles and lower energies, Solar magnetic fields and Earth
magnetic field asymmetries may alter the expectation beyond
statistical uncertainties.  In Fig.~\ref{fig:FERMIsensitivity}, we
plot the differential flux for several benchmark signals, namely one- and two-step cascade decays for $1\TeV$ and $2\TeV$ DM mass. The fluxes are normalized so that the FSR differential photon flux at $10\GeV$ is
$d\Phi_\gamma /dE = 3\times 10^{-6}\rm{/m^2/s/GeV}$, hence saturating the last bin of the Fermi solar result~\cite{Giglietto}. The fluxes for the different benchmarks are given in Table~\ref{tbl:fluxes}. Fig.~\ref{fig:FERMIsensitivity} also includes background estimates with statistical uncertainty
for a few choices of $\Delta\theta$. Given the figures for the different models, Solar
electronic fluxes as small as $F\sim 10^{-4}~{\rm m}^{-2}{\rm
  s}^{-1}$ at energies $E\gsim 100 \GeV$ should be detectable with
existing Fermi data. 

\begin{table}[htdp]
\caption{Fluxes of the annihilation products from the Sun used in Fig.~\ref{fig:FERMIsensitivity}. The fluxes are normalized so that the FSR differential photon flux at $10\GeV$ is
$d\Phi_\gamma /dE = 3\times 10^{-6}\rm{/m^2/s/GeV}$ saturating the last bin of the Fermi solar result~\cite{Giglietto}. To obtain the actual electronic flux these figures should be multiplied by 2 (4) for the one- (two-) step cascades, assuming the annihilation products decay entirely into $e^+e^-$ pairs.}
\begin{center}
\begin{tabular}{|c|c|}
\hline
Model & Flux ($\rm{m^{-2}s^{-1}}$)
\\
\hline
one-step, $1\TeV$ & $2.7\times 10^{-4}$\\
one-step  $2\TeV$ & $2.5\times 10^{-4}$\\
two-step  $1\TeV$ & $1.6\times 10^{-4}$\\
two-step  $2\TeV$ & $1.4\times 10^{-4}$\\
\hline
\end{tabular}
\end{center}
\label{tbl:fluxes}
\end{table}%

Such an approach helps with the defocusing effects of the Solar magnetic fields on the electron fluxes, but what about the possibilities of a coherent deflection from the Parker spirals? A simple approach for this would be to look for ``hotspots'' in the solar frame. That is, one can consider the flux $\frac{d\Phi(\theta)}{dE}$ in regions of size $\theta$ which tile the sky. In observing this distribution, there will certainly be variation from place to place, but the appearance of outliers (i.e., anomalously large $\frac{d\Phi(\theta)}{dE}$ in one direction compared to the others) could give evidence of solar production, even if this hotspot were not centered on the sun. The expectation is that while some deflection is possible, extremely large deflections ($\gsim 30^\circ$) are unlikely. Although this technique is less sensitive, it overcomes the complications associated with a coherent deflection of the electronic flux. As a final corollary to this point, we note that while {\em discoveries} can be made by looking at smaller regions around the sun, one should be careful in placing {\em limits} from looking at regions smaller than about $30^\circ$. Should the signal arrive from a different angle, highly focused regions could give anomalously strong limits on electronic production.

\section{Conclusions}\label{sec:Conclusion}

In this paper we proposed two novel mechanisms by which high energy electrons originating from DM annihilations in the Sun can arrive at the Earth's vicinity. Such high energy electron flux should be observable already with existing Fermi data and may help resolve the outstanding discrepancy with the ATIC results. If such a flux is observed, it will constitute an unambiguous evidence for DM annihilations in the Sun. It will also force a revision in our ideas about the way DM interacts with matter and/or the nature of its annihilation products. 

\begin{acknowledgments}
We thank J. Wefel for useful discussions of the ATIC experiment, and
S. Murgia and N. Giglietto for useful discussions of the Fermi solar
gamma-ray measurements.  PS is supported by the US DOE under contract
number DE-AC02-76SF00515. NW is supported by NSF CAREER grant
PHY-0449818 and DOE OJI grant \# DE-FG02-06ER41417.  IY is supported
by the James Arthur Fellowship.

\end{acknowledgments}

\bibliography{SolarPaper}
\end{document}